\documentclass[fleqn,usenatbib]{mnras}

\usepackage[T1]{fontenc}

\DeclareRobustCommand{\VAN}[3]{#2}
\let\VANthebibliography\thebibliography
\def\thebibliography{\DeclareRobustCommand{\VAN}[3]{##3}\VANthebibliography}

\usepackage{graphicx}	
\usepackage{amsmath}	
\usepackage{amssymb}	
\usepackage{soul}	

\title[The 6/1 MMR between Quaoar's ring and Weywot]{Dynamical characterization of the 6/1 mean motion resonance between Quaoar's ring and Weywot}

\author[A. Rodr\'iguez et al.]{
Adri\'an Rodr\'iguez,$^{1}$\thanks{E-mail: adrian@ov.ufrj.br}
B. E. Morgado,$^{1}$ 
N. Callegari Jr.$^{2}$
\\
$^{1}$Observat\'orio do Valongo, Universidade Federal do Rio de Janeiro, Ladeira do Pedro Ant\^ onio 43, 20080-090, Rio de Janeiro, Brazil\\
$^{2}$S\~ao Paulo State University (UNESP), Institute of Geosciences and Exact Sciences, Av. 24-A, 1515, 13506-900, Rio Claro, SP, Brazil
}

\begin{document}
\label{firstpage}
\pagerange{\pageref{firstpage}--\pageref{lastpage}}
\maketitle


\begin{abstract}
Recently, it has been reported the discovery of a dense ring around the trans-Neptunian object 50000 Quaoar. The ring particles seem to be very close to the 6/1 mean motion resonance with Weywot, the only known satellite in the system. In this work we investigate the dynamical environment in the close vicinity of the 6/1 orbital resonance in the context of the restricted three body problem. We aim to analyze whether, in view of observational constraints, the ring could be effectively evolving in resonant motion with the satellite. Through the technique of dynamical maps we identify and characterize the 6/1 mean motion resonance, finding that the main location of the resonance deviates by only $29$ km from the central part of the ring. This difference lies within the 3$\sigma$ confidence level, considering the uncertainties in the observational parameters. We also show that the Weywot's eccentricity plays a significant role in the dynamical structure of the 6/1 resonance. The results show that the resonance width is smaller than the estimated ring's width. Under assumption of a ring with eccentricity smaller than 0.05, clumping of test particles appears at the position of the different resonant multiplets, considering the nominal value of Weywot's eccentricity. This is in agreement with observations, which indicate that the estimated resonance width ($\leq$ 10 km) is comparable with the narrow and dense arc of material within Quaoar's ring. Our results may be an indicative that the 6/1 resonance resonance plays a key role in confining the arc ring.

\end{abstract}




\section{Introduction}
\label{intro}

\onecolumn


Rings of particles around small bodies seem more common than previously thought. The first example of its kind was discovered around the Centaur 10199 Chariklo \citep{Braga-Ribas-2013} in 2013. In 2017, a somewhat similar structure was observed around the dwarf planet 136108 Haumea \citep{Ortiz_2017} and more recently around the trans-Neptunian object (TNO) 50000 Quaoar \citep{Morgado_2023}. Besides those, a ring was also proposed around the Centaur 2060 Chiron \citep{Ortiz_2015, Sickafoose_2020}, but it is yet to be confirmed by better observational data.

Among those rings, Quaoar's ring has a unique property: it is far outside the expected Roche Limit \citep{Morgado_2023}. In fact, classical models suggest that particles orbiting at a distance $r$ from a body with mass $M$ should have densities larger than the Roche critical density ($\rho_{cr}$) to not accrete into a satellite. Based on this, \cite{Morgado_2023} suggests that Quaoar's Roche limit would be near 1,780 km, assuming particles with densities of 400 kg m$^{-3}$ (typical of the inner Saturnian moons). This value is much smaller than the ring radius, which was estimated in $\sim$4,100 km by \cite{Morgado_2023} and recently improved in $\sim$4,057 km by \cite{Pereira_2023}, taking into account new observational data. Particles with densities larger than this threshold would rapidly accrete into a moon within a few decades \citep{Kokubo_2000, Takeda_2001}.

Another intriguing feature of Quaoar's ring is its azimuthal inhomogeneity and irregularity, from dense and narrow parts to wider and less dense regions. With this, Quaoar's ring bear a resemblance with Saturn's F ring \citep[][and references therein]{Murray_2018} and Neptune's Adams rings \citep{Hubbard_1986}. Nonetheless, the Quaoar's ring optical depth indicates that collisions have a key role in the dynamics of this ring.

Dynamical mechanisms are needed to explain the observational characteristics of Quaoar's ring. One key aspect is the mean motion resonance (MMR) between the ring particles and Quaoar's moon, Weywot. In fact, as suggested by \cite{Morgado_2023}, the ring is close to the 6/1 MMR with Weywot, located about 4,030 km away from Quaoar center. However, uncertainties in the observational parameters could move this resonance as far as a hundred kilometers. A preliminary theoretical analysis of the 6/1 MMR is presented by \cite{Morgado_2023}, thus, without performing numerical simulations in a representative phase space.  

In this paper, we evaluate the dynamical stability in the vicinity of Quaoar's ring using numerical integration of the exact equations of motion in the context of the restricted three body problem. Our goal is to identify whether the ring (or some part of it) is currently evolving in resonant motion with Weywot.

In Sec. \ref{model} we present the adopted model for numerical simulations. The main results are shown in Sec. \ref{results}, where we investigate the dynamical environment in the ring's close vicinity (\ref{results1}), the dynamical structure of the 6/1 MMR (\ref{results2}) and the analysis in the space of frequencies (\ref{results4}). Discussion and conclusions are devoted to Sec. \ref{conclusion}.

\section{Model and methods}         
\label{model}

We investigate the dynamical environment of the ring particles and the 6/1 MMR with Weywot through dynamical maps. This technique is widely used in order to identify regular and chaotic motions in the space of orbital elements of perturbed bodies \citep[e.g.][]{michtchenko&ferrazmello2001,Callegari+2021}. It is also very useful for identification and characterization of MMRs. The dynamical maps are constructed through a grid of numerical simulations of the exact equations of motions in the framework of the restricted (non circular) three body problem. We consider Quaoar as a central body, Weywot as the perturbing satellite and a massless particle. We also include the contribution of the Quaoar's oblateness parameterized by the $J_2$ coefficient of the gravitational potential\footnote{The zonal coefficient $J_2$ of a triaxial body can be computed based on the mass of the body ($M_Q$), the equivalent area radius ($R$), and its principal moments of inertia \citep[][and references therein]{Hu_2004}. The principal moments of inertia ($I_{\sf xx}$, $I_{\sf yy}$, and $I_{\sf zz}$) can be computed based on the ellipsoid semi-axis ($a$, $b$, and $c$) following the classical formalism by \cite{Murray_1999}. Assuming Quaoar is a Maclaurin spheroid as proposed by \cite{Braga-Ribas-2014} with axis $a~=~b~=569$ km and $c~=~519$ km, then $J_2$ can be estimated as $J_2=0.0354$. It is important to highlight that this value is an estimation based on the nominal parameters presented in \cite{Braga-Ribas-2014}. Future observations can provide better values for this parameter.}. We use MERCURY in order to perform the numerical simulations \citep{chambers1999}, choosing the Bulirsch-Stoer algorithm. The integration time is 200 yr or 1,000 yr, depending on specific maps, and the step size is taken as 0.1 days. We adopt 150$\times150=22,500$ as a size grid, varying the initial semimajor axis and eccentricity of particles and also the mass of Weywot. For each dynamical map we calculate the maximum variation of test particles orbital eccentricities ($\Delta e=e_{max}-e_{min}$) as dynamical indicator. According to observations \citep{Morgado_2023}, the preferred ring's orbit is aligned with the orbit of Weywot, hence, we neglect all inclinations in the simulations. In addition, the initial argument of pericenters, ascending nodes and mean anomalies are assumed to be zero. Table \ref{tabela-system} displays physical parameters and orbital elements of the Quaoar - Weywot system. The ring properties are summarized in \cite{Morgado_2023}.

\begin{table}
\begin{center}
\caption{Physical parameters and orbital elements adopted in the numerical simulations. The data is taken from \citep{Morgado_2023} and the reader can obtain further details, including errors bars, in the cited work. 
The physical and orbital parameters of Quaoar and Weywot were obtained from the method described in \citep{Vachier+2012}.The radius of Weywot was taken from \citep{Kretlow2020}. The mass of Weywot was obtained through the radius and assuming that its mean density equals the Quaoar's mean density (1.99 gcm$^{-3}$ \citep{Braga-Ribas-2014}).}
\vspace*{0.3cm}
\begin{tabular}{c | c c c c c}
\hline
\hline
Body & Mass (kg) & Radius (km) & $J_2$ & $a$ (km) & $e$ \\ 
\hline
\textbf{Quaoar} & 1.20$\times10^{21}$ & 555 & 0.0354 & - & - \\
\textbf{Weywot} & 5.12$\times10^{18}$ & 85 & 0 & 13,289 & 0.056 \\
\hline
\label{tabela-system}
\end{tabular}
\end{center}
\end{table}

\section{Results}
\label{results}

\subsection{Dynamical environment in the ring's close vicinity} 
\label{results1}

\begin{figure}
\begin{center}
\includegraphics[width=0.6\columnwidth,angle=270]{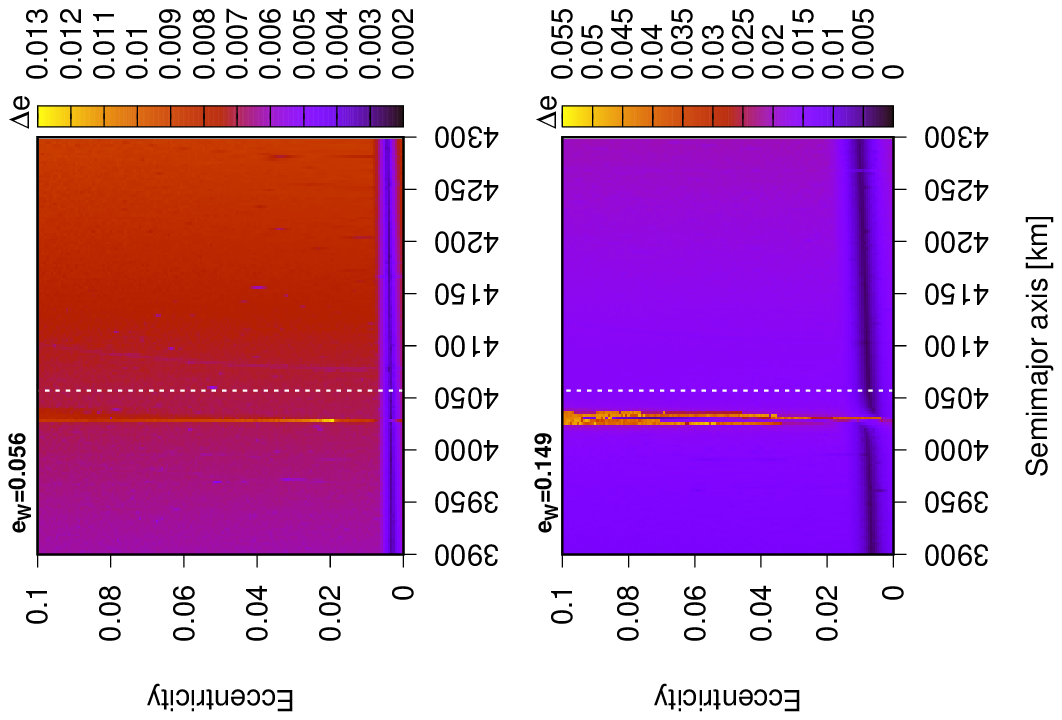}
\caption{Dynamical maps for test particles in the close vicinity of the Quaoar's ring. The grid contains 22,500 numerical simulations of the exact equations of motion integrated by 1,000 yr. We show the maximum variation of the test particles eccentricity as dynamical indicator. Two dynamical maps are shown, one for the nominal (top panel) and other for the maximum (bottom panel) value of the Weywot's eccentricity, according to observational constraints \citep[see][]{Morgado_2023, Pereira_2023}. The dotted vertical line indicates the solution for the nominal radius of the ring, at 4,057 km.}
\label{mapas-full}
\end{center}
\end{figure}

Firstly, we perform a set of dynamical maps in the space of semimajor axis and eccentricity of test particles. We adopt three different values of the Weywot's eccentricity ($e_W$), since this orbital quantity has a strong uncertainty ($e_W=0.056\pm0.093$, according to \cite{Morgado_2023}). Fig. \ref{mapas-full} shows two dynamical maps in the space of semimajor axis and eccentricity for $e_W=0.056$ (nominal value) and $e_W=0.0149$ (maximum value).
All 22,500 numerical simulation were integrated by 1,000 yr ($\sim 30,000$ Weywot's orbital periods). The dotted vertical line indicates the solution for the nominal radius of the ring, at 4,057~km.

On one hand, we identify the location of the 6/1 MMR with Weywot closely at $a_{6/1}=4,028$ km in both dynamical maps of Fig. \ref{mapas-full}. As expected, the domain and strength of the resonance increase with $e_W$. At first hand, it appears that the ring location is well separated from the position of the 6/1 MMR resonance by 29 km. However, considering the uncertainties on Quaoar's mass, Weywot semi-major axis and ring parameters, we can not eliminate the possibility of this ring to lie within this resonance. As expected, the dynamical environment for $e_W=0.056$ is smoother than for $e_W=0.149$. In the first   and latter case, the maximum variation of eccentricity corresponds to $\Delta e=0.013$ and $\Delta e=0.055$, respectively, attained at a semimajor axis $\simeq a_{6/1}$.

\begin{figure}
\begin{center}
\includegraphics[width=0.35\columnwidth,angle=270]{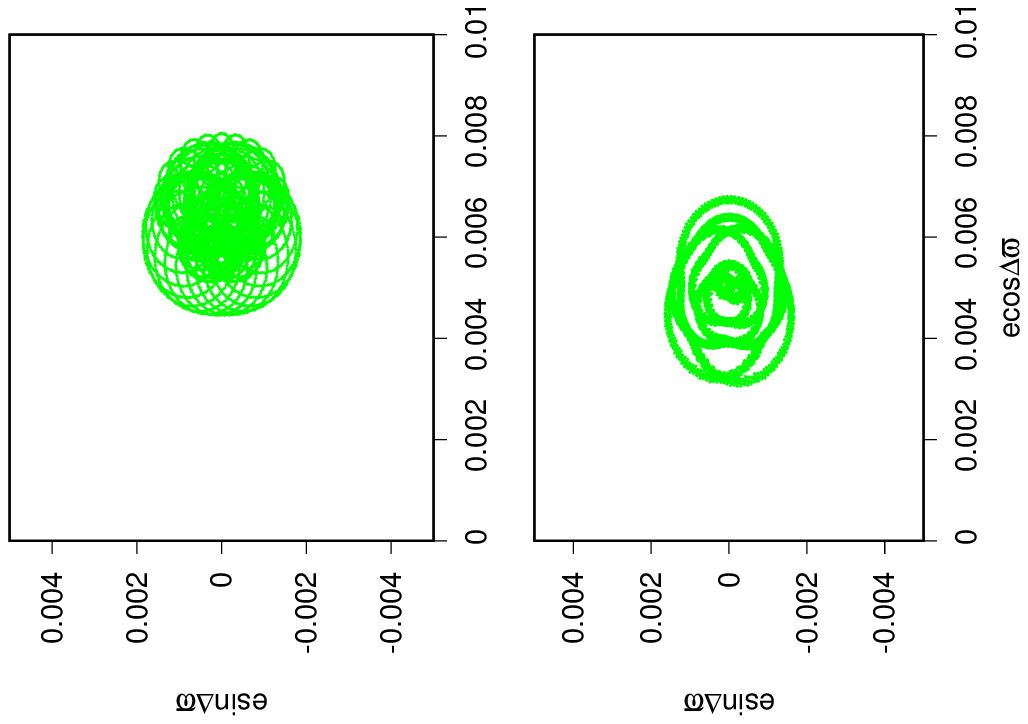}
\caption{The variation of $\Delta\varpi=\varpi_W-\varpi$ in the representative plane $(e\cos\Delta\varpi,e\sin\Delta\varpi)$ corresponding to two individual simulations with the following initial conditions: $a=4,102$ km, $e=0.00805$ and $a=3,925$ km, $e=0.00668$. In both cases, $e_W=0.149$ and all angles are initially equal to zero.}
\label{dpi}
\end{center}
\end{figure}

It is interesting to note in both maps of Fig. \ref{mapas-full} the raising of a closely horizontal structure with very small values of $\Delta e$. We performed two individual simulations corresponding to the white dots in the bottom panel of Fig. \ref{mapas-full}. The results are shown in Fig. \ref{dpi}, where we can see the oscillation of $\Delta\varpi=\varpi_W-\varpi$ around 0 (hereafter, angles without sub-index correspond to test particles and "W" stand for Weywot's angles). Therefore, we conclude that the structure appearing for small eccentricity corresponds to the place of equilibrium points of the secular angle $\Delta\varpi$.

\begin{figure}
\begin{center}
\includegraphics[width=0.6\columnwidth,angle=270]{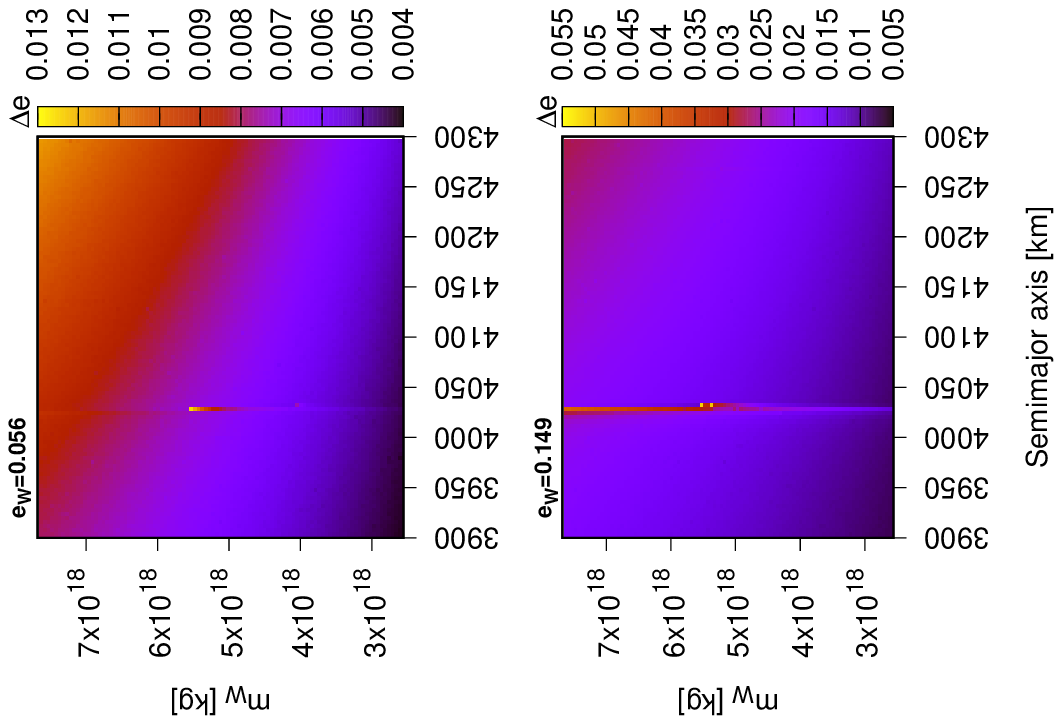}
\caption{The same than Fig. \ref{mapas-full} here varying the mass of Weywot and fixing the initial test particles eccentricities at $10^{-4}$.}
\label{mapas-full-massa}
\end{center}
\end{figure}

For sake of completeness, we perform a new set of dynamical maps now varying the Weywot's mass ($m_W$), keeping fixed the value of the initial eccentricities of test particles at $10^{-4}$. We adopt a range of mass such that $0.5\leq \rho_W/\rho_Q\leq1.5$, where  $\rho_Q$ and $\rho_W$ are the mean densities of Quaoar and Weywot, respectively, resulting in $2.6\times10^{18}$ kg $\leq m_W\leq7.7\times10^{18}$ kg. The results are shown in Fig.  \ref{mapas-full-massa}, for the same previously adopted values of $e_W$ and integration time. The signature of the 6/1 MMR appears in both dynamical maps, with strong significance for larger $e_W$. Moreover, larger semimajor axes and Weywot's masses result in the highest values of $\Delta e$. It is important to note that, despite the width of the resonance increases for large $m_W$, the value of $a_{6/1}$ remains unchanged, as can be seen in Fig.  \ref{mapas-full-massa}.

All numerical simulation concerning Figs. \ref{mapas-full} and \ref{mapas-full-massa} shown that test particles survives the whole integration time in stable motion. The collected above results allows us to conclude that the long-term dynamical evolution of test particles in the close vicinity of the ring is very regular, with no evidence of chaotic or irregular motion. In the next section we will explore the detailed structure of the 6/1 MMR, including the motion of test particles in the resonant domain as well in the separatrix.

It is important to highlight, according to our numerical simulations, the absence of additional structures other than the 6/1 MMR and the region corresponding to the oscillation of $\Delta\varpi$. 



\subsection{Dynamical structure of the 6/1 MMR}
\label{results2}

We perform a new set of dynamical maps in the space of initial semimajor axis and eccentricity of test particles, focusing in the close vicinity of the 6/1 MMR with Weywot. The aim is the characterization and identification of resonant structures. Moreover, we intend to investigate how the resonant domain depends on the eccentricity of the satellite. We adopt four values of $e_W$, namely, 0.028, 0.056, 0.1 and 0.149, covering the most part of the uncertainty in the eccentricity.

\begin{figure}
\begin{center}
\includegraphics[width=0.6\columnwidth,angle=270]{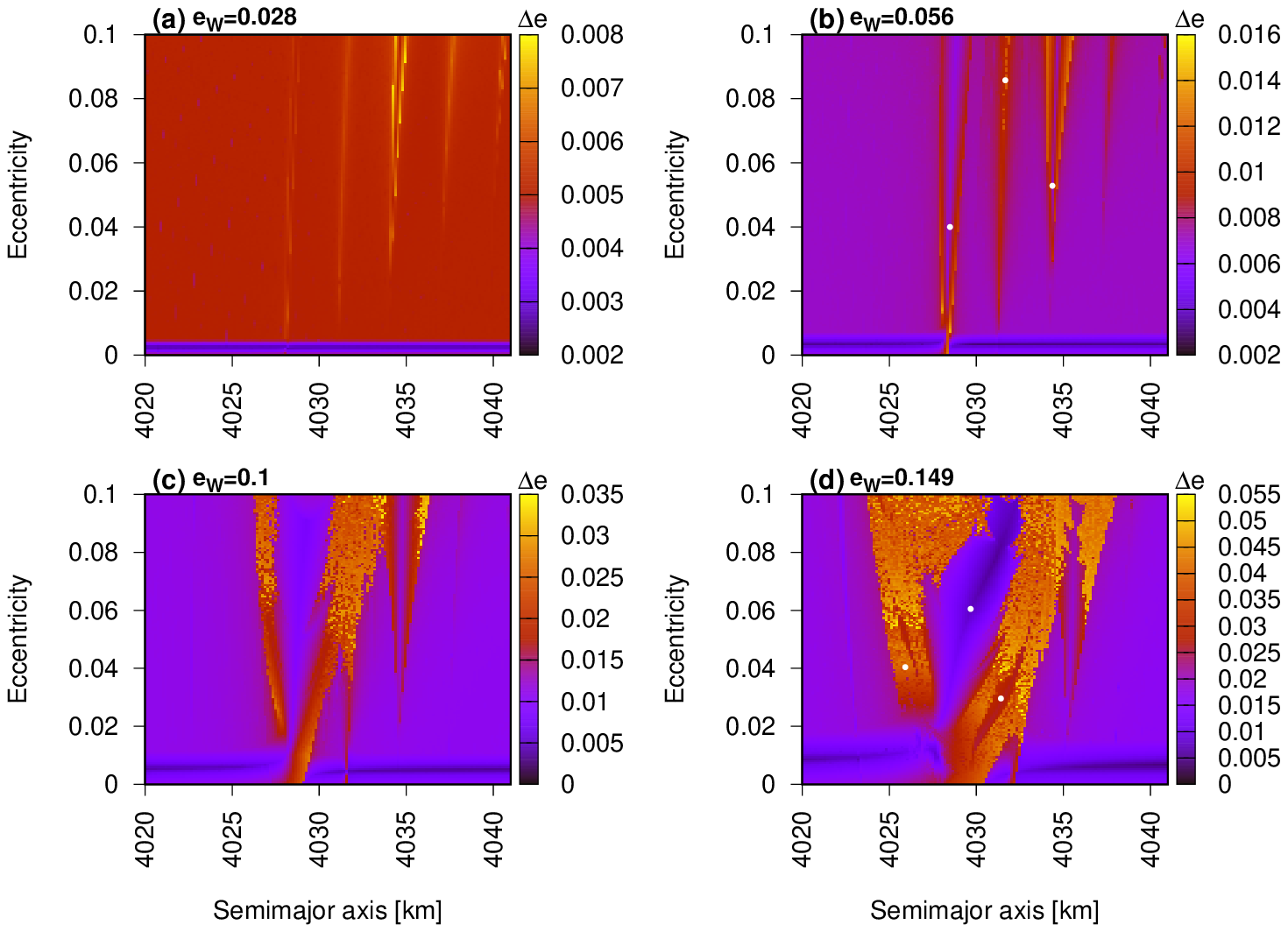}
\caption{Dynamical maps for four values of the Weywot's eccentricity within the observational constraints. The integration time of each of 22,500 numerical simulation is 200 yr and all initial angles are equal to zero. As in the previous maps, we plot here $\Delta e $ as dynamical indicator. The white dots indicate the initial conditions used for individual simulations (see Sec. \ref{results3}).}
\label{mapas-zoom}
\end{center}
\end{figure}

Fig. \ref{mapas-zoom} shows the dynamical maps for the adopted values of $e_W$. The integration time for these maps was 200 yr, covering roughly several times the period of the resonance (depending on the particle's eccentricity; see more details in Secs. \ref{results3} and \ref{results4}). Panels (a) and (b) show the case for $e_W=0.028$ and $e_W=0.056$, respectively. We note the raising of three main resonant signatures at 4,028 km, 4,031 km, and 4,034 km, more evident in the case $e_W=0.056$. This feature is a natural consequence of resonant systems evolving under $J_2$ perturbations and it is known as resonance splitting \citep[see][]{Peale1999}. It is important to note that these structures correspond to librations of three different critical angles of the 6/1 MMR (see Sec. \ref{results3}). Indeed, we have that $\phi_1=6\lambda_W-\lambda-\varpi-4\varpi_W$, $\phi_2=6\lambda_W-\lambda-2\varpi-3\varpi_W$ and $\phi_3=6\lambda_W-\lambda-3\varpi-2\varpi_W$ librates around 0, as it is shown in Fig. \ref{ecce-sigma-nom}. The other three multiplets associated to the 6/1 commensurability are $\phi_0=6\lambda_W-\lambda-5\varpi_W$, $\phi_4=6\lambda_W-\lambda-4\varpi_W-\varpi$ and $\phi_5=6\lambda_W-\lambda-5\varpi$ \citep[see][]{Morgado_2023}. They do not appear in our dynamical maps because they are very weak, however, we return to this discussion in Sec. \ref{results4}.

Panels (c) and (d) show the results for the cases $e_W=0.1$ and $e_W=0.149$. As the eccentricity of the satellite increases, the domain of the principal resonance (the one at 4,028 km) overlaps with the other multiplets, only appearing the one at 4,034 km. According to expectations, the width of the resonance increases for high $e_W$. Moreover, $\Delta e$ also increases, mainly in those regions close to the separatrix.

The results shown in Fig. \ref{mapas-zoom} allows us to claim that, depending on the Weywot's eccentricity, the 6/1 MMR structure varies from a set of multiplets with $\sim$1 km of width to an almost single component with $\sim$10 km of width. These values are in agreement with the dense arc of material within Quaoar's ring, which may be a indicative that this resonance plays a significant role in confine the arc.

\subsubsection{Individual orbits}
\label{results3}

\begin{figure}
\begin{center}
\includegraphics[width=0.4\columnwidth,angle=270]{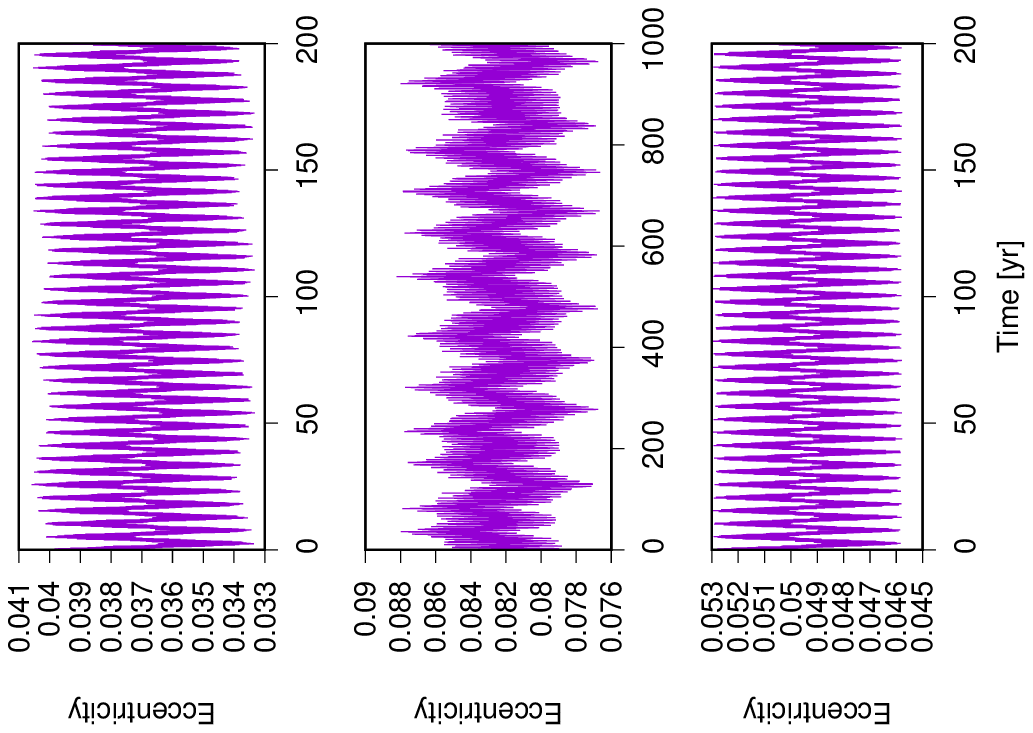}
\includegraphics[width=0.4\columnwidth,angle=270]{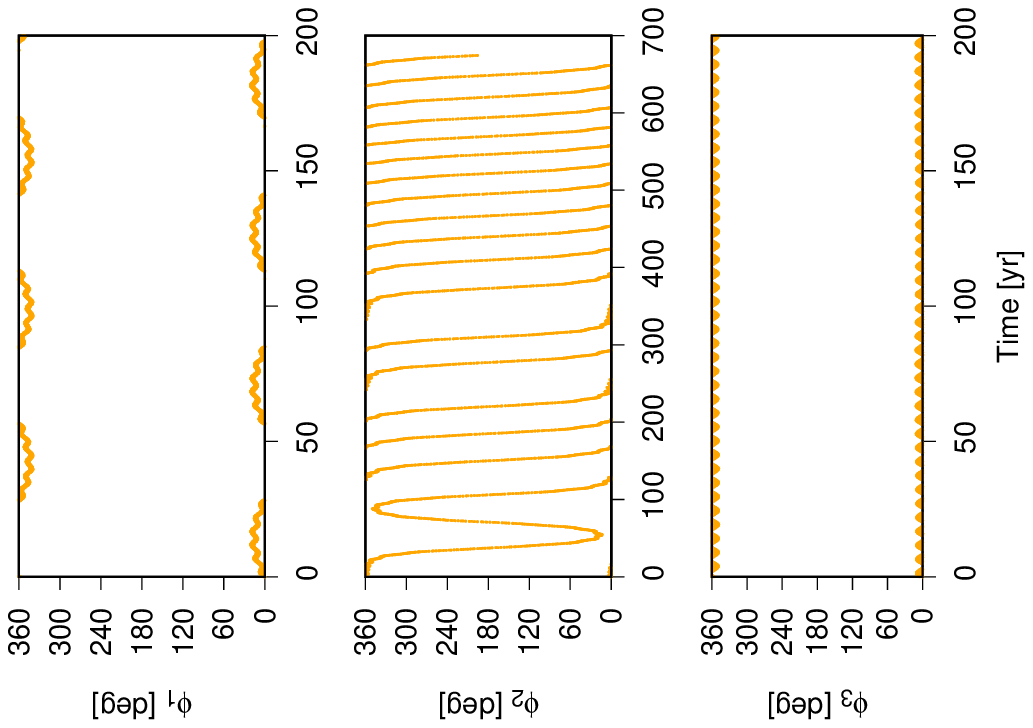}

\caption{Time variation of test particles eccentricity (left) and critical angle of the 6/1 MMR (right). The initial conditions correspond to the white dots in Fig. \ref{mapas-zoom}(b), namely, $a=4,027$ km, $e=0.0375$ (top), $a=4,031$ km, $e=0.0857$ (centre) and $a=4,034$ km, $e=0.0529$ (bottom). The angles $\phi_1=6\lambda_W-\lambda-\varpi-4\varpi_W$ and $\phi_3=6\lambda_W-\lambda-3\varpi-2\varpi_W$ librates around 0, whereas $\phi_2=6\lambda_W-\lambda-2\varpi-3\varpi_W$ alternates between circulation and episodic libration periods. In the three cases, $e_W=0.056$ and all angles are initially equal to zero.}
\label{ecce-sigma-nom}
\end{center}
\end{figure}

We numerically integrate a few particular cases in order to better analyze the motion of test particles near the center, border and close vicinity of the 6/1 MMR. The time variation of test particles eccentricities and critical angles are shown in Fig. \ref{ecce-sigma-nom} for the case $e_W=0.056$ and for initial conditions indicated by white dots in Fig. \ref{mapas-zoom}(b). The angles $\phi_1$ and $\phi_3$ ($\phi_2$) librates with very small (high) amplitude around 0\footnote{Note that $\phi_2$ also shows some circulations.}. These results allows us to identify the different components of the 6/1 MMR appearing for the nominal value of Weywot's eccentricity.

\begin{figure}
\begin{center}
\includegraphics[width=0.4\columnwidth,angle=270]{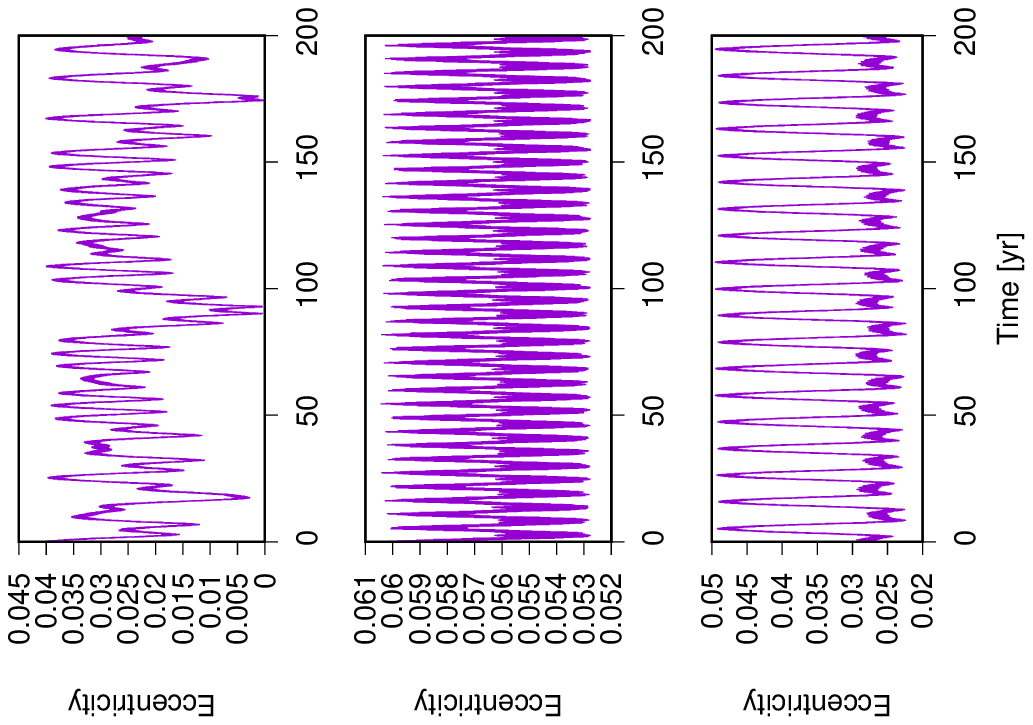}
\includegraphics[width=0.4\columnwidth,angle=270]{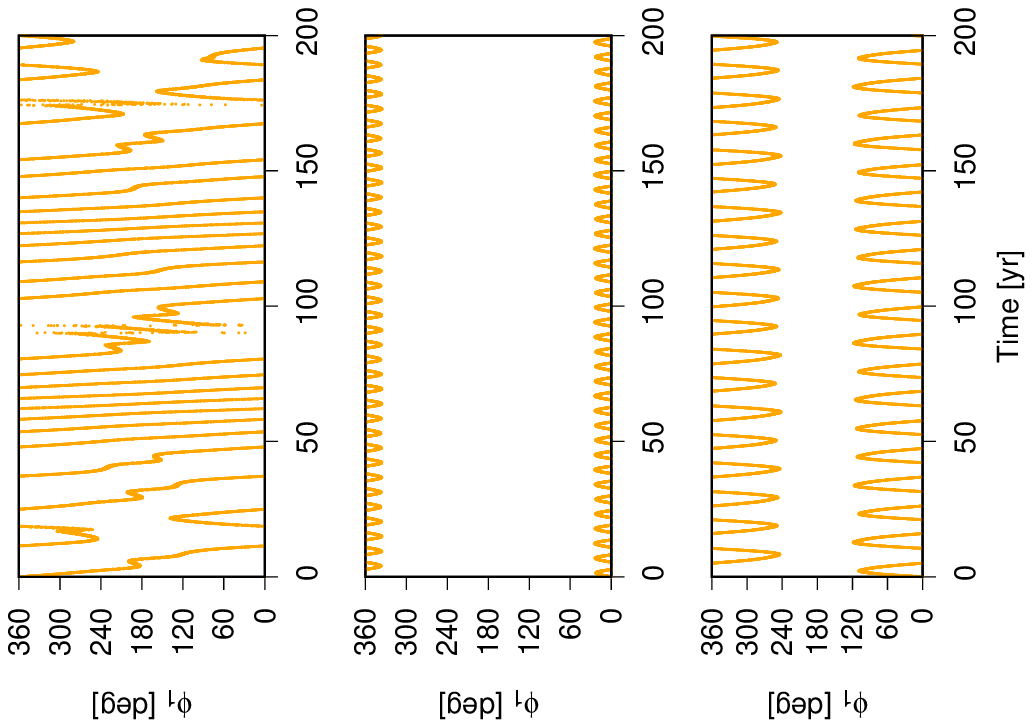}

\caption{The same than Fig. \ref{ecce-sigma-nom}, here for $e_W=0.149$. We only show the angle corresponding to the multiplet centred at $\sim 4,028$ km, that is, $\phi_1$. The initial conditions are $a=4,026$ km, $e=0.0375$ (top), $a=4,030$ km, $e=0.0605$ (centre) and $a=4,032$ km, $e=0.0296$ (bottom), indicated by white dots in Fig. \ref{mapas-zoom}(d).}
\label{ecce-sigma-max}
\end{center}
\end{figure}

Fig. \ref{ecce-sigma-max} shows the case for $e_W=0.149$ and represented by the initial conditions corresponding to white dots in Fig. \ref{mapas-zoom}(d). Note that these points are located close to the center and separatrix of the resonance. The initial condition close to the center (middle panels) evolves with very small amplitude of eccentricity, whereas the critical angle $\phi_1$ librates around 0. The top and bottom panels show the initial condition at the left and right borders of the V-shaped resonance structure, respectively, according to Fig. \ref{mapas-zoom}(d). The initial condition at the left shows $\phi_1$ in a regime of circulation, with episodic cycles of libration around 0, which is a typical behavior of motion evolving close to the resonance separatrix. Note that the condition starting at the right border evolves under libration of $\phi_1$, with amplitude $\sim 120^{\circ}$.

We stress that all critical angles corresponding to initial conditions of Figs. \ref{ecce-sigma-nom} and \ref{ecce-sigma-max} result in circulation of $\Delta\varpi$.

\subsection{Frequency analysis}
\label{results4}

Fig. \ref{mapas-zoom} shows the complexity of the domains of the 6/1 MMR and its dependence with the orbital eccentricity of Weywot, which is poorly constrained. Now, let us investigate in more details the web of resonances associated to the 6/1 commensurability. For this task, we apply the technique of frequency analysis of a dense set of orbits of test particles numerically integrated \citep[see][and references therein]{Callegari+2021}. We make an unidirectional sweeping of the resonance by considering the initial semi-major axis ($a_0$) of test particle as the free parameter in the simulations. The results are the ``individual dynamical power spectra'', denoted by IPS. The y-axes of the IPSs give, for each initial condition, the periods associated to the peaks in the spectrum of some variable of the problem which have amplitudes larger than a prefixed fraction which we define by reference amplitude, RA. In general RA is of the order of a few percent, and the variable is the semi-major axis or the orbital eccentricity of particles. The frequencies are obtained with the Fast Fourier Transform method \citep[e.g.][]{Press1996}. It is worth to note that only the initial semi-major axes are changed, so that all other initial elements of the particles and Weywot are always fixed. As in the dynamical maps, all initial angles are assumed to be zero.

\begin{figure}
\centerline{\includegraphics[width=0.9\columnwidth]{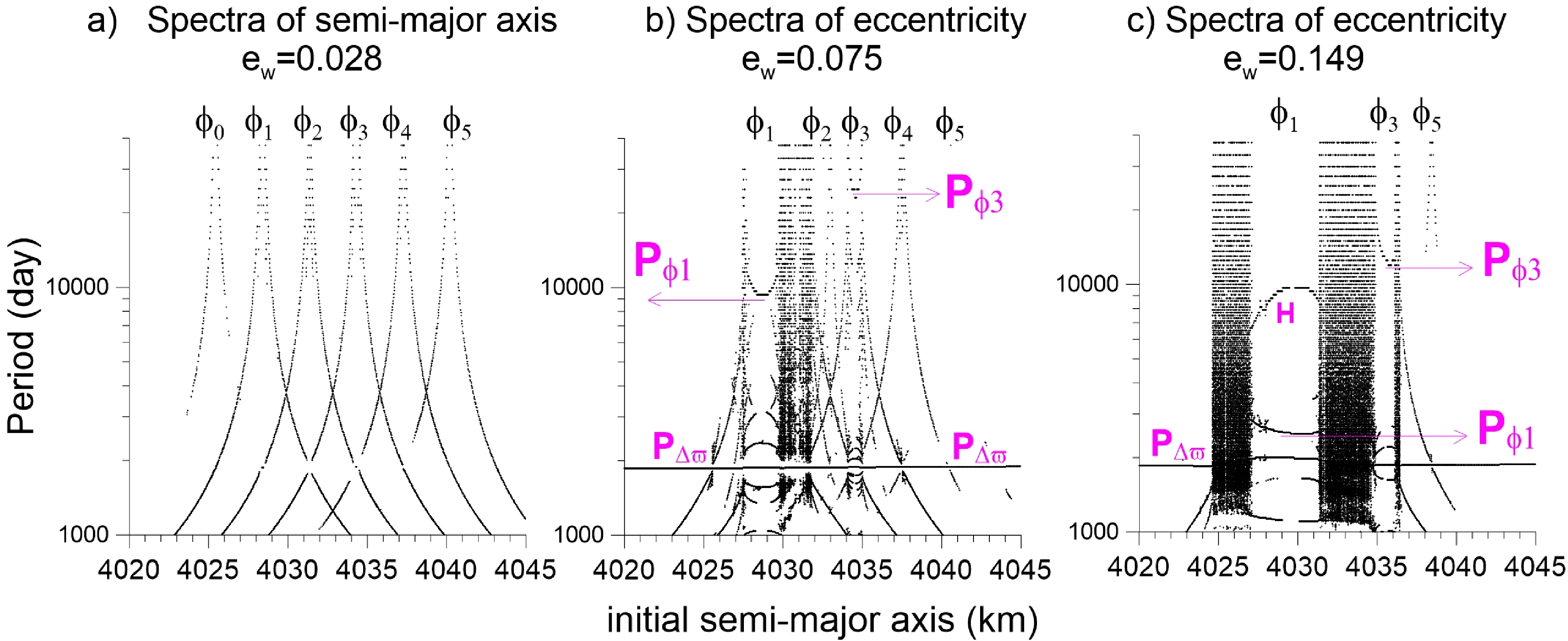}}
\caption{Individual dynamical power spectra of 1500 test particles orbiting Quaoar disturbed by Weywot and the $J_2$ of Quaoar. Different variable have been utilized in order to calculate the spectra, as well distinct values of the eccentricity of Weywot, $e_W$. See mais text of Section 3.3 for details. The periods associated to the fundamental frequencies are indicated by arrows. $\phi_k$ are the multiplets of the 6/1 particle-Weywot resonance (see definitions in Section 3.1). The orbital eccentricity of all particles is $0.06$.}
\label{<ips>}
\end{figure}

Fig. \ref{<ips>} shows three IPSs obtained for three different values of $e_W$, namely, $e_W=0.028$, $e_W=0.075$ and $e_W=0.149$. They have been obtained from the spectra of semi-major axis (left), eccentricity (middle and right) of 1500 test particles in the range $4,020\leq a_0\leq 4,045$ km. The initial eccentricity of the particles is $0.06$. RA is $5\%$ (left) and $1\%$ (middle and right), and y-axes are given in logarithmic scale.

Fig. \ref{<ips>}a shows the loci of all multiplets associated to the 6/1 resonance. Their definitions are given in Sec. \ref{results2}. For $e_W=0.028$, the centre of the $\phi_0$, $\phi_1$, $\phi_2$, $\phi_3$, $\phi_4$, $\phi_5$ resonances are given at $a_0\sim4,025.4$ km, $a_0\sim4,028.4$ km, $a_0\sim4,031.3$ km, $a_0\sim4,037.2$ km and $a_0\sim4,040.2$ km, respectively.

As $e_W$ increases, the nominal positions of the resonances suffer small changes of the order of a few km or less (see Figs. \ref{<ips>}b,c). However, two main features arise for larger value of $e_W$: the increasing of the volume of some resonances in the phase space, and the merging of other ones leading to overlap. In fact, Fig. \ref{<ips>}a reveals that the domains of the $\phi_1$-multiplet is larger, in such a way that we can identify the loci of the period associated to the resonance, \textbf{$P_{\phi_1}$}. For $e_W=0.075$,  \textbf{$P_{\phi_1}\sim26$} yr. Thus, there is a smooth variation of \textbf{$P_{\phi_1}$} with $a_0$, such that in the narrow range of $a_0$ centered at $a_0\sim4,029$ km, the dominant peak in the spectra of the eccentricity of the particles is the one associated to the $\phi_1$ resonance.

The continuation of \textbf{$P_{\phi_1}$} is interrupted at the left and right borders of the resonance, where two vertical ``barriers'' are reached. The spectra of orbits have a large number of peaks surrounding the fundamental frequencies, showing their irregular (and possibly chaotic) nature. This occurs since these regions of the phase space are located within the separatrices of the resonance \citep[see][]{Callegari+2021}. Moreover, still considering Fig. \ref{<ips>}b, we see that the domains of $\phi_1$ and $\phi_2$ resonances merge at $a_0\sim4,031$ km leading to chaotic orbits around their separatrices (given by vertical barriers).

Increasing $e_W$ at the largest allowed value, $e_W=0.149$, it is possible to note the overlap of $\phi_2$ resonance with the separatrices of the $\phi_1$ and $\phi_3$ multiplets (Fig. \ref{<ips>}c). At such large value of $e_W$, the domains of the $\phi_1$ and $\phi_3$ resonances increase significantly (see Fig. \ref{mapas-full-massa}d), and their associated periods are now one order of magnitudes smaller than previously, such that \textbf{$P_{\phi_1}\sim7$} yr and \textbf{$P_{\phi_3}\sim2.5$} yr, respectively.

The isolated (almost) horizontal line in Figs. \ref{<ips>}b,c close to $\sim 2000$ day ($\sim 5$ yr) is the functional dependence of the period of circulation of $\Delta\varpi$ with $a_0$, and it is indicated by \textbf{$P_{\Delta\varpi}$} (see Section 3.1 for definition). Note that for $e_W=0.075$ the continuation of the horizontal line structure is not broken at the serapatrices (Fig. \ref{<ips>}b), but for larger $e_W$, the same doesn't occur (Fig. \ref{<ips>}c).

Finally, we explain the line indicated by \textbf{$H$} in Fig. \ref{<ips>}c. It corresponds to a multiplet involving the fundamental frequencies \textbf{$P_{\phi_1}$}  and  \textbf{$P_{\Delta\varpi}$} such that \textbf{$P_{H}=P_{(\Delta\varpi-\phi_1)}=1/P_{\Delta\varpi}-1/P_{\phi_1}\sim27$ } yr.

\subsection{Searching for arcs ring}
\label{results5}

\begin{figure}
\begin{center}
\includegraphics[width=0.5\columnwidth,angle=270]{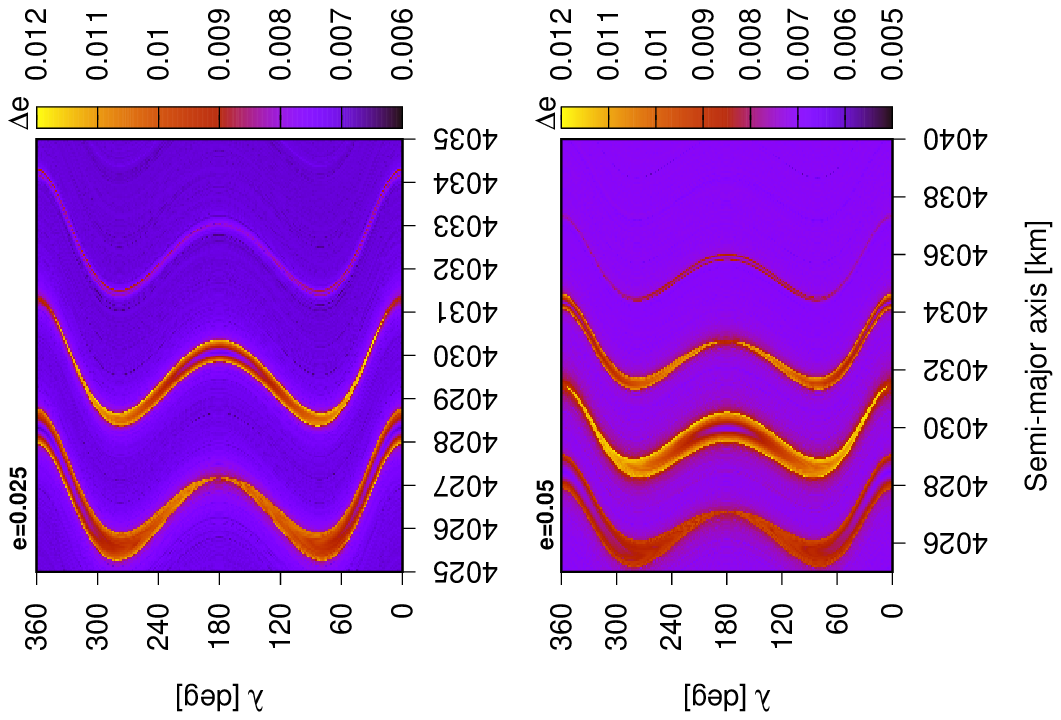}
\includegraphics[width=0.5\columnwidth,angle=270]{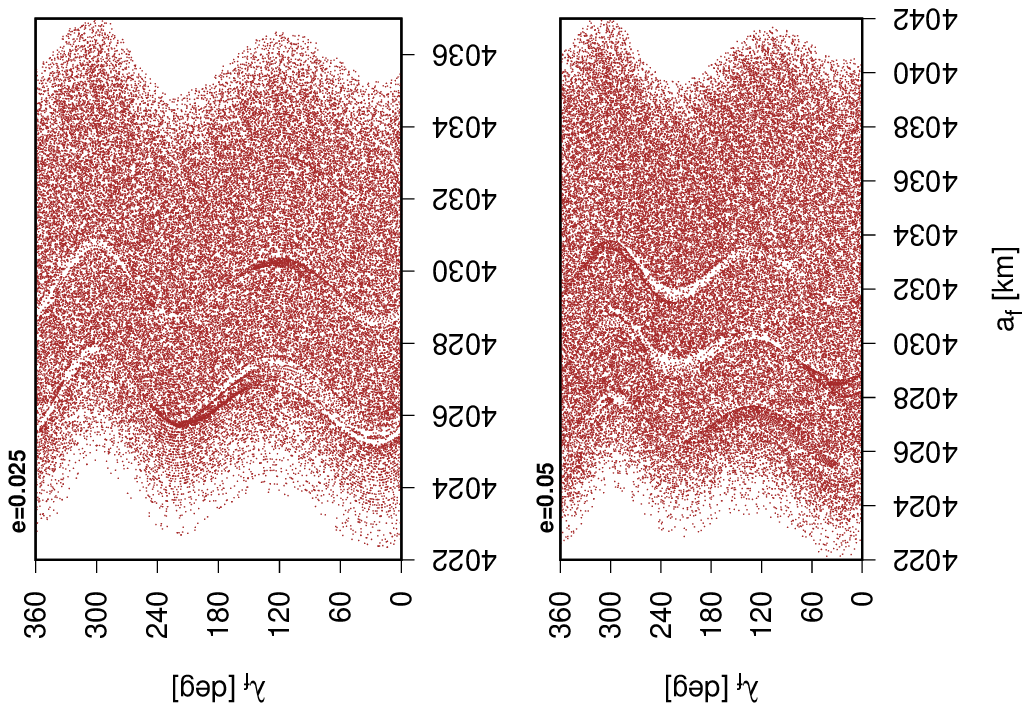}

\caption{\textit{Left column}: dynamical maps varying the initial semi-major axis and mean longitude of test particles (see text for details). The integration time is 200 yr and we consider two values of test particles eccentricity, namely, $e=0.025$ (top) and $e=0.05$ (bottom). \textit{Right column}: Final distribution of semi-major axis and mean longitudes of test particles, searching for high density locations of particles, assuming the same two previous initial values of $e$.}
\label{arcos}
\end{center}
\end{figure}

As mentioned before, Quaoar's ring is highly irregular, and present dense regions of material that can be interpreted as clumps or arc of material akin to Saturn's F ring and Neptune's Adams rings \citep{Morgado_2023, Pereira_2023}. In view of the observational evidence for the presence of arcs ring, we perform new dynamical maps now varying the initial mean longitude of test particles, keeping fix their initial eccentricities. A similar analysis was performed by \cite{rodriguez+callegari2021} in the context of small satellites arcs (Aegaeon, Methone and Anthe). We choose $e=10^{-4}$, $e=0.025$ and $e=0.05$ as initial values of particles orbital eccentricities. Fig. \ref{arcos} shows the results for $e=0.025$ and $e=0.05$, plotting the dynamical maps (left column) and also the final values of mean longitudes and semi-major axis, $\lambda_f$ and $a_f$ (right column). The initial angles other than mean anomalies for particles are chosen to be zero and the simulation time is 200 yr, as in the case of previous dynamical maps. In all simulations, $e_W=0.056$ and the initial angles of Weywot's orbit are zero. 

On one hand, we identify in the dynamical maps the multiplets corresponding to the case of $e_W=0.056$ (see Fig. \ref{mapas-zoom}b). Note as the multiplets are highlighted when $e$ increases. Taking into account the definition of $\phi_k$ and the initial conditions for the angles, we have $\lambda=\phi_k$ (for all $k$) in the maps. It is interesting to note the rising of the librations centers (0 and $180^{\circ}$) for the different $\phi_k$. On the other hand we note, in the plot of final semi-major axis and mean longitudes, several places of clumping with large density of particles when compared with the close vicinity. As expected, these clumps are located at the positions of the different $\phi_k$ corresponding to the 6/1 MMR with Weywot. For $e=0.025$, we note two clumps corresponding to $\phi_1$ and $\phi_2$. As the domain of $\phi_3$ is very week at $e=0.025$ (see Fig. \ref{mapas-zoom}b and left panel of Fig. \ref{arcos}), we do not find clumps at $\phi_3$. In the case of $e=0.05$, we have three clumps corresponding to $\phi_1$, $\phi_2$ and $\phi_3$, where now the domain of $\phi_3$ is more evident. Moreover, these clumps seem to be less dense than for the case of $e=0.025$. It is important to emphasize that the case for $e=10^{-4}$ have not shown any evidence of clumping, indicating that the arcs only appear for a non circular ring.  

The above results are in concordance with ring's observational constraints. In fact, assuming that the ring can be described as a group of nested elliptical streamlines associated with an azimuthal harmonic number $m~=~1$, as a first approximation, we can estimate a lower limit to Quaoar's ring eccentricity based on the ring's width variation ($\Delta W_r$) and ring radius ($r$) \citep[][and references therein]{French_1986, Berard_2017, Morgado_2021}. Using the values in \cite{Morgado_2023}, the changes in the ring eccentricity ($\delta e$) can be computed by $\delta e = \Delta W_r/2r$, using $\Delta W_r\simeq295$ km and $r\simeq4,057$ km, resulting in $\delta e\simeq0.036$. By analogy with Uranus and Saturn ringlets, the ring eccentricity ($e$) is expected to be slightly larger than $\delta e$ \citep{French_1986}.

\section{Conclusions and discussion}
\label{conclusion}

We have investigated the dynamical environment in the close vicinity of the Quaoar's ring in the framework of the restricted three body problem. The dynamical maps, constructed through a grid of thousands of numerical simulations of the exact equations of motion, revealed a regular behavior with small values of maximum variations of semi-major axis and eccentricity (Figs. \ref{mapas-full} and \ref{mapas-full-massa}). In addition, we identified the location of the 6/1 MMR between the ring and the satellite Weywot, placed at about 4,028 km. We do not detected additional mean motion resonances with Weywot that may affect the dynamics of the ring.

According to observational constraints, the Weywot's eccentricity can vary between 0 and 0.149 \citep{Morgado_2023}. We have shown that $e_W$ plays a fundamental role in determine the dynamical structure of the 6/1 MMRs with ring particles. For $e_W\lesssim0.075$, the 6/1 MMR divides into several multiplets whose widths are of the order of 1 km and are separated by a few km, occupying the range between 4,028 km and 4,037 km. Among these multiplets, the dominants ones are those whose resonant angles are given by $\phi_1=6\lambda_W-\lambda-\varpi-4\varpi_W$ and $\phi_3=6\lambda_W-\lambda-3\varpi-2\varpi_W$, located at around 4,028 km and 4,034 km. For $e_W\gtrsim0.1$ the multiplets corresponding to $\phi_1$ becomes the dominant, with a width of $\sim10$ km, overlapping with the $\phi_3$ multiplet for eccentricity $\gtrsim0.06$ (Fig. \ref{mapas-zoom}). We also determined the period of the main multiplet ($\phi_1$), which varies from a few years to a few decades, depending on $e_W$ ($P_{\phi_1}$ decreases with $e_W$).  In addition, we have found that ring particles with $e<0.01$ may be secularly evolving with pericenters aligned with the Weywot's pericenter (Figs. \ref{mapas-full} and \ref{mapas-full-massa}). Further observational data may provide constraints in the Weywot's eccentricity, allowing for an accurate characterization of the 6/1 MMR with the ring.

We have also found that, under assumption of a ring with eccentricity of the order of a few per cent, clumping of test particles appear at the position of the different multiplets of the 6/1 MMR, considering the nominal value of Weywot's eccentricity. This is in agreement with observations, which point that the Quaoar's ring has regions of large density of material that can be interpreted as arcs ring, indicating that the 6/1 MMR between Weywot and the ring's particles can play an important role in confining arcs within this ring.

A single resonance between the ring and a perturber may cause the migration of the ring material due to the action of torques. \citep{Sicardy+1987,Sicardy+2019}. The torque ($\Gamma$) applied to the ring due to the perturbation of Weywot can be estimated by using Eq. (3) of \cite{Sicardy+2019}. The timescale for the dissipation of the ring due to the torque is thus given by $\tau=J/\dot{J}=J/\Gamma$, with $J$ being the orbital angular momentum of the ring particles.
After replacing numerical values, we obtain $\tau\sim10^{12}$ yr for the nominal Weywot's eccentricity. The above results indicate that, neglecting dissipative forces, the torque is unable to dissipate the ring in reasonable timescales\footnote{We note that the torque is an upper limit of the torque applied to the ring, as only part of the ring may be located on the resonance region. In addition, the torque is valid for an specific multiplet associated to the Lindblad resonance, namely, $\phi_1$. We have shown that the 6/1 MMR displays regions of irregular motion when $e_W$ is large, making the calculation of the torque meaningless in this case. We also note that $\tau$ estimates the timescale over which the ring is significantly radially displaced due to the torque. An additional relevant timescale is the time to radially displace the ring by the width of the resonance (few km, depending on $e_W$), that would be orders of magnitude smaller that the previous estimate.}.

Within the limitations of our model we remark: i) the absence of additional perturbations such as 
the Sun, dwarf planets and non-conservative forces like radiation pressure, etc \citep[see][]{araujo+2016}. ii) The non consideration of mutual interaction between ring particles. iii) The poor constraint in the shape of Quaoar, making its $J_2$ value to be highly uncertain. Current and upcoming missions may provide additional observational data for the Quaoar's system, allowing for a better characterization of the 6/1 MMR between the ring and Weywot.


\section*{Acknowledgements}

We acknowledge the anonymous reviewer for his/her revision and for the helpful suggestions, which allowed us to improve the manuscript. We thank to Bruno Sicardy for valuable comments and suggestions. AR is grateful to FAPERJ (process 210.419/2022). NCJ is grateful to FAPESP (processes 2020/06807-7). BEM is grateful to CNPq (process  150612/2020-6).

\section*{Data Availability}

The data underlying this article will be shared on reasonable request to the corresponding author.



\bibliographystyle{mnras}
\bibliography{accepted} 



\bsp	
\label{lastpage}
\end{document}